\documentclass[preprint,showpacs,preprintnumbers,amsmath,amssymb]{revtex4}
\usepackage{graphicx}
\usepackage{dcolumn}
\usepackage{bm}

\newcommand{\be}{\begin{equation}}
\newcommand{\en}{\end{equation}}
\newcommand{\bea}{\begin{eqnarray}}
\newcommand{\ena}{\end{eqnarray}}

\begin{document}


\title{ Reconstructing  warm inflation  }

\author{Ram\'on Herrera}
\email{ramon.herrera@pucv.cl} \affiliation{ Instituto de
F\'{\i}sica, Pontificia Universidad Cat\'{o}lica de
Valpara\'{\i}so, Casilla 4059, Valpara\'{\i}so, Chile.}

\date{\today}

\begin{abstract}
The reconstruction of a warm inflationary universe model from the
scalar spectral index   $n_S(N)$ and the tensor to scalar ratio
$r(N)$ as a function of the number of e-folds $N$ is studied.
Under a general formalism we find the effective potential and the
dissipative coefficient in terms of the cosmological parameters
$n_S$ and $r$ considering  the weak and strong dissipative stages
under the slow roll approximation. As a specific example, we study
the attractors for the index $n_S$ given by $n_{S}-1\propto
N^{-1}$ and for the ratio $r\propto N^{-2}$, in order to
reconstruct the model of warm inflation. Here, expressions for the
effective potential $V(\phi)$ and the dissipation coefficient
$\Gamma(\phi)$ are obtained.

\end{abstract}

\maketitle

\section{Introduction}

It is well known that during
 the evolution of the early universe, it exhibited  an accelerated  expansion
 or an
inflationary scenario commonly  called the inflationary
universe\cite{I1,Staro}. A crucial characteristic of the
inflationary universe is that this scenario explicates  the
Large-Scale Structure (LSS) of the universe, and also the source
of the anisotropies observed in the Cosmic Microwave Background
(CMB) radiation\cite{I3}. Although,  inflation originally was
proposed to solve some problems of the standard hot bing-bang
model such as; the flatness, horizon,  among other\cite{I1,Staro}.

In the context of  the different models that give account of the
inflationary universe and its early evolution, we can distinguish
the model of warm inflation. In  the framework  of warm inflation,
the universe is described  by a self-interacting
  radiation field and a field scalar or inflaton field.
  In contradiction to the standard cold inflation, the model of
  warm inflation has the
attractive feature that it avoids the reheating period, because
the radiation production takes place  concurrently together with
the inflationary expansion driven by the scalar field \cite{warm,1126,warm2}. This is possible through
 a friction term enclosed on the dynamical equations  and this term  describes
the processes of the scalar field dissipating into a thermal bath
 with other fields. In this sense, the
scenario of  warm inflation ends whenever the universe stops
inflating and softly  goes into the radiation epoch of the
standard big-bang model.

Another difference of warm inflation  in relation to the cold
inflation are the initial fluctuations  essential  for the LSS
formation.  In fact,  during the development   of warm inflation
the thermal fluctuations have a fundamental  role in the LSS
formation and the density fluctuations from the scalar field arise
from thermal rather than quantum fluctuations
\cite{62526,Herrera:2017qux}. Thus, from the background dynamics  and
initial fluctuations, the stage of warm inflation differs
substantially from the cold inflation (or the standard
inflation)\cite{taylorberera}. For a review of models of warm
inflation, see e.g. Refs.\cite{w1,taylorberera,w2,warm2} and for a
list of recent articles, see \cite{w3,w4}.

On the other hand, the reconstruction of the  effective potential
 in the evolution of cold inflation from
observational data such as the scalar spectrum,  scalar spectral
index $n_S$ and the tensor to scalar ratio $r$,   have been
discussed by several authors\cite{H1,H2,H3,H4,M,Chiba:2015zpa,H5}.
An interesting  mechanism in order to construct the effective
potential of inflation assuming the slow roll approximation, is
through the parametrization of the cosmological parameters or attractors
$n_S(N)$ and $r(N)$, where $N$ corresponds to the number of
e-folds. The observational tests from Planck data\cite{Planck} are
in good accord with the parametrization on the  scalar spectral
index given by $n_S\sim 1-2/N$ and the tensor to scalar ratio
$r\propto N^{-2}$, assuming that the number $N\simeq 50-70$ at the
end of the inflationary epoch. For large $N$ ($N\gg 1$) the
attractor $n_S(N)\sim 1-2/N$ together with different expressions
for the tensor to scalar ratio $r(N)$ can be deduced from
different models in the case of cold inflation such as; the
T-model \cite{T}, E-model\cite{E}, Staronbisky
$R^2$-model\cite{Staro}, the chaotic model\cite{Linde83}, the
model of Higgs inflation with non minimal
coupling\cite{Higgs,Higgs2} among other.

On the other hand, it is also possible to consider the slow-roll
parameter $\epsilon$ and its parametrization in terms of $N$, in
order  to obtain the effective potential, scalar spectral index
and the tensor to scalar ratio in models of   cold inflation
\cite{Huang:2007qz,M,Gao:2017owg}. In particular in Ref.\cite{M}
was studied different types of slow-roll parameter $\epsilon(N)$
and thereby reconstructing  the effective potential. Also, from
the two slow roll parameters $\epsilon(N)$ and $\eta(N)$
 the
effective potential was reconstructed  in
Ref.\cite{Roest:2013fha}. Analogously,   in Refs.\cite{N1, N2}
related results are obtained for the reconstruction.

The objective of this article is to reconstruct the model of the
warm inflation, considering the parametrization  of the
cosmological parameters  as the scalar spectral index and the
tensor to scalar ratio in terms of the number of e-folds. In this
context, we analyze how the background dynamics in which  there is
a self-interacting scalar field and radiation affects the
reconstruction of the effective potential and the dissipative
coefficient from the attractors. Under a general formalism, we
will build the potential and dissipative coefficient during the
scenario of the weak and strong dissipative regimes from the
attractors $n_S(N)$ and $r(N)$. Also, for the reconstruction the
model of warm inflation we will consider  the weak and strong
dissipative regimes assuming  the slow roll approximation.

In order to reconstruct analytical quantities for the potential
and dissipative coefficient, we will study a concrete example for
the cosmological parameters  $n_S(N)$ and $r(N)$. Here, we will
consider the attractors $n_s-1\propto 1/N$ and $r\propto 1/N^2$,
for two regimes during the stages of  warm inflation. In both
scenarios, we will find the potential and dissipative coefficient
together with the constraints on the different parameters assuming
the condition for the weak regime and  the strong regime,
respectively.

The outline of the article is a follows: the next section presents
a short review of  the basic equations during the stage of warm
inflation.  In the section \ref{secti2a} we discuss the
reconstruction in the framework of  warm inflation. In sections
\ref{section3a} and   \ref{section3b} we obtain  under a general
formalism, explicit expressions for the  effective potential and
dissipative coefficient in terms of the number of e-folds $N$
during
 the weak and strong dissipative regimens, respectively. In section
\ref{example} we discuss  a concrete example for our model, in
which we consider the specific  attractors for $n_S(N)$ and
$r(N)$, in order to find the potential $V(\phi)$ and the
coefficient $\Gamma(\phi)$.
 Here, we analyze the reconstruction in the
weak and strong dissipative regimes, respectively.
 Finally, our conclusions are presented in
Section\ref{conclu}.  We chose units so that $c=\hbar=8\pi G=1$.

\section{Warm inflation:basic relations\label{secti} }

We start by writing down  the  Friedmann equation in the framework
of the warm inflation, by considering a spatially flat Friedmann
Robertson Walker (FRW) metric, together  with a scalar field
homogeneous and  radiation. In this sense, the Friedmann equation
is given by

\begin{equation}
H^2=\frac{1}{3}\,\rho=\frac{1}{3}\,[\rho_{\phi}+\rho_\gamma]\,,
\label{HC}
\end{equation}
where $H=\dot{a}/a$ denotes the Hubble parameter and the quantity
$a$ corresponds to the scale factor. During the scenario of warm
inflation we assume a two-component system, a scalar field
homogeneous $\phi=\phi(t)$ with an energy density $\rho_\phi$ and
a radiation field of energy density $\rho_\gamma$. Here, the total
energy density $\rho=\rho_\phi+\rho_\gamma$, where the energy
density $\rho_\phi$ in terms of the scalar field is defined  by
$\rho_\phi=\dot{\phi}^2/2+V$, where  $V$ denotes the effective
potential. In the following, we will consider that the dots
correspond to differentiation with respect to the time.

As previously mentioned in the framework of the warm inflation, the universe is filled
with a self-interacting scalar field and radiation, and
 the basic equations for the densities $\rho_\phi$ and
$\rho_\gamma$ are given by \cite{warm}
 \be\dot{\rho_\phi}+3\,H\,(\rho_\phi+p_\phi)=-\Gamma\;\;\dot{\phi}^2,\,\,\,\mbox{or equivalently}\,\,\,
 \ddot{\phi}+[3H+\Gamma]\dot{\phi}=-\frac{\partial V}{\partial\phi}=-V_{,\phi}\,\;, \label{key_01}
 \en
and
 \be \dot{\rho}_\gamma+4H\rho_\gamma=\Gamma\dot{\phi}^2
.\label{3}\en  Here, we mention that the continuity equation for
the total energy density $\rho$ satisfies the standard relation
$\dot{\rho}+3H(\rho+p)=0$.

In this context, the quantity $\Gamma$ refers to the dissipation
coefficient and considering  the second law of thermodynamics the
 coefficient $\Gamma$ is defined as
positive\cite{warm,62526}. In this sense, from Eqs.(\ref{key_01})
and (\ref{3}) we interpret that the coefficient $\Gamma$ gives
origin to
 the decay of the scalar field
into radiation during the inflationary epoch of the universe. The
parameter $\Gamma$ can be considered to be a constant or a
function of the scalar field $\phi$, or the temperature of the
thermal bath $T$, or both i.e., $\Gamma=\Gamma(\phi,T)$
\cite{warm}.

In the following we will analyze the reconstruction of the model
of warm inflation, assuming that the dissipation coefficient and
the effective potential depend only of the scalar field, i.e.,
$\Gamma=\Gamma(\phi)$ and $V=V(\phi)$, respectively.

During the inflationary epoch, the energy density of the scalar
field predominates over the energy density associated to the
radiation field in warm inflation, wherewith
$\rho_\phi>\rho_\gamma$. Also, considering the set of slow-roll
approximations in which $\dot{\phi}^2\ll V$ and
$\ddot{\phi}\ll(3H+\Gamma)\dot{\phi}$,  then the Friedmann
equation (\ref{HC}) can be rewritten as\cite{warm,62526}
\begin{eqnarray}
H^2\simeq\frac{1}{3}\,\rho_\phi\simeq \frac{1}{3}\,V,\label{inf2}
\end{eqnarray}
and from  the equation of the scalar field (\ref{key_01})  we have
\begin{equation}
 \dot{\phi}\simeq -\frac{V_{,\phi}}{3H(1+R)}.\label{inf3}
\end{equation}
Here, the quantity $R$ corresponds to the ratio between the
coefficient $\Gamma$ and the Hubble parameter  and is defined as
\begin{equation}
 R=\frac{\Gamma}{3H }.\label{rG}
\end{equation}
Typically, during   the scenario of  warm inflation we can
identify two regimes; called
  the  weak  dissipative
regime, in which  $R\ll 1$ (or equivalently $\Gamma\ll3H$) and the
strong dissipative regime where the ratio $R\gg 1$
($\Gamma\gg3H$).

On the other hand, following Refs.\cite{warm,62526}, we  assume
that during the inflationary expansion of the universe   the
radiation production is quasi-stable, in which
$\dot{\rho}_\gamma\ll 4 H\rho_\gamma$ and $
\dot{\rho}_\gamma\ll\Gamma\dot{\phi}^2$. In this form,  from
Eq.(\ref{3}) we find that the energy density of the radiation
field becomes
 \begin{equation}
\rho_\gamma= C_\gamma\,
T^4\simeq\frac{\Gamma\dot{\phi}^2}{4H}=\frac{R}{4(1+R)^2}\,\frac{V_{,\phi}^2}{V},\label{rh}
\end{equation}
 where $C_\gamma$ is a constant and is defined as
$C_\gamma=\pi^2\,g_*/30$, in which $g_*$ denotes the number of
relativistic degrees of freedom\cite{warm,warm2}.  Here, we have
used Eq.(\ref{inf3}).

Also,  from Eq.(\ref{rh}), we find that temperature of the thermal
bath can be written as
\begin{equation}
T= \left[\frac{R}{4C_\gamma\,(1+R)^2}\,\frac{V_{,\phi}^2}{V}\right]^{1/4}.\label{T}
\end{equation}
In order to have a measure of the inflationary expansion of the
universe, we define the number of e-folding $N$ between two
different values of cosmological times $t$ and $t_e$, where the
time $t_e$ corresponds to the end of inflation. Thus, the number
of e-folds $N$ assuming the slow roll approximation can be written
as
\begin{equation}
N=\int_t^{t_e}H\,dt'=\int_\phi^{\phi_e}H\,\frac{d\phi'}{\dot{\phi}}\simeq\int_{\phi_e}^\phi
\,\frac{V(1+R)}{V_{,\phi}}d\phi'.\label{3N}
\end{equation}
Here, we have considered Eqs.(\ref{inf2}) and (\ref{inf3}).

On the other hand, due to the presence of the radiation field in
the dynamics of warm inflation,  the source of the density
fluctuations correspond to thermal
fluctuations\cite{warm,1126,warm2}. In this sense, during the
evolution of the expansion inflationary,  the fluctuations of the
scalar field  are dominantly thermal rather than
quantum\cite{warm,62526,Herrera:2017qux}. Thus, in the scenario of
warm inflation, the curvature and entropy perturbations coexist,
since the mixture of the scalar field and radiation are generate
at the perturbative levels. This happens because the model of warm
inflation can be viewed as a model of two basic fields\cite{Jora}.
However, as was demonstrated in Ref.\cite{warm2}, during warm
inflation the entropy perturbations on the large scales decay  and
only the curvature
 (adiabatic modes) survives\cite{warm,warm2,62526,Herrera:2017qux}. In this context,
the power spectrum of the curvature perturbations ${\cal{P_S}}$
during warm inflation, assuming  $\Gamma=\Gamma(\phi)$ and
$V=V(\phi)$ together with the slow roll approximations becomes
 \cite{warm2,fNL,Moss:2008yb,culia}
\begin{equation}
{\cal{P_S}}\simeq\,\frac{H^3\,T}{\dot{\phi}^2}\,\sqrt{(1+R)}.
\end{equation}

 The scalar spectral index $n_S$ is defined as $n_S=d\ln{\cal{P_S}}/d\ln k$ and  in terms of the
slow roll parameters is given by\cite{Moss:2008yb}
\begin{equation}
n_S-1=-\frac{(9R+17)}{4(1+R)^2}\,\epsilon-\frac{(9R+1)}
{4(1+R)^2}\,\beta+\frac{3}{2}\frac{1}{(1+R)}\,\eta,\label{nsG}
\end{equation}
where the slow roll parameters $\epsilon$, $\eta$ and $\beta$ are
given by
\begin{equation}
\epsilon=\frac{1}{2}\left(\frac{V_{,\phi}}{V}\right)^2,\;\,\,\eta=\frac{V_{,\phi\phi}}{V},
\,\,\,\mbox{and}\,\,\,\beta=\frac{V_{,\phi}\,\Gamma_{,\phi}}{V\Gamma}.
\end{equation}
Here, we mention that in the case in which $\Gamma=\Gamma(\phi,T)$ and 
$V=V(\phi,T)$, we should add two new parameters in Eq.(\ref{nsG}) for the scalar spectral index $n_S$, see 
Ref.\cite{Moss:2008yb}.

On the other hand,  since the tensor perturbation do not couple to the thermal
background, then this suggests that
 the tensor modes have an equivalent amplitude to the model of cold
inflation, where the tensor spectrum ${\cal{P_T}}$ is defined as
${\cal{P_T}}=8\,H^2$. Thus, the tensor to scalar ratio $r$ in the
scenario of warm inflation can be written as
\begin{equation}
  r=\frac{{\cal{P_T}}}{{\cal{P_S}}}=\frac{8\dot{\phi}^2}{HT\sqrt{1+R}}=
  \frac{16\epsilon}{(1+R)^{5/2}}\frac{H}{T}.\label{gr}
\end{equation}
Here, we have used Eq.(\ref{inf3}).
We note that the tensor to scalar ratio $r$ in the model of warm
inflation, see Eq.(\ref{gr}), cannot be written only in terms of
the slow-roll parameter $\epsilon$ as it occurs in cold inflation,
in which $r=16\epsilon$. Also, we observe that Eq.(\ref{gr}) coincides with the 
ratio $r$ 
obtained in Ref.\cite{culia}.

In the following we will study the reconstruction of the effective
potential $V$ and the dissipative coefficient $\Gamma$ in the
scenario
 of
warm inflation, considering an attractor point from  scalar spectral index $n_S(N)$ and
the tensor to scalar ratio $r(N)$ in the $r-n_S$ plane.

\section{reconstruction  \label{secti2a} }

In this section and following Ref.\cite{Chiba:2015zpa},
 we explicate the procedure  to follow in the reconstruction of the
effective potential and the dissipation coefficient as a function
of the scalar field $\phi$ in the framework of warm inflation,
considering  the scalar spectral index $n_S(N)$ and the tensor to
scalar ratio $r(N)$ as attractors. Since we have two quantities
$V(\phi)$ and $\Gamma(\phi)$ in the reconstruction, we need first
of all to express  the scalar spectral index and the tensor to
scalar ratio in terms of the number of e-folds $N$. For this it is
necessary to rewrite Eqs.(\ref{nsG}) and (\ref{gr})  in terms  of
the potential and the dissipation coefficient as a function of the
number of e-folds $N$ and its  derivatives. Thus, from these
equations and giving $n_S=n_S(N)$ and $r=r(N)$  we should obtain
the effective potential and the dissipation coefficient as a
function of the number $N$. Posteriorly, from Eq.(\ref{3N}) we
should find the e-folds $N$ in terms of the scalar field in order
to reconstruct the potential $V(\phi)$ and the coefficient
$\Gamma(\phi)$, respectively.

 We start by rewriting
 the index and ratio given by  Eqs.(\ref{nsG})
and (\ref{gr}) in terms of the number of e-folding $N$. In fact,
the slow roll parameters can be rewritten in terms of the number
of e-folds $N$, considering that
$$
V_{,\phi}=\frac{dV}{d\phi}=\frac{V(1+R)}{V_{,\phi}}\,V_{,\,N}\,,
$$
then we get
\begin{equation}
V_{,\phi}^2=V(1+R)\,V_{,\,N}\,,\,\,\,\,\mbox{wherewith}\,\,\,\,V_{,\,N}=\frac{dV}{dN}>0.\label{dV}
\end{equation}
In the following, we will consider the subscript $V_{,\,N}=dV/dN$,
$V_{,NN}$ to $V_{,NN}=d^2V/dN^2$, $\Gamma_{,\,N}=d\Gamma/dN$ etc.

Analogously for $V_{,\phi\phi}$ we have
\begin{equation}
V_{,\phi\phi}=\frac{1}{2V_{,\,N}}\left[(1+R)[V_{,\,N}^2+V\,V_{,\,NN}]+V\,V_{,\,N}\,R_{,\,N}
      \right],\label{ddV}
\end{equation}
and for the quantity $\Gamma_{,\phi}$ we have
\begin{equation}
\Gamma_{,\phi}=\left[\frac{V(1+R)}{V_{,\,N}}\right]^{1/2}\,\Gamma_{,\,N}.
\end{equation}

Thus, the slow roll parameters
can be rewritten as
\begin{equation}
\epsilon=\frac{V_{,\,N}}{2V}\,(1+R),\;\;\;\;\,\,\beta=\frac{\Gamma_{,\,N}}{\Gamma}\,(1+R),\label{p1}
\end{equation}
and
\begin{equation}
\eta=\frac{1}{2VV_{,\,N}}\left[(1+R)[V_{,\,N}^2+V\,V_{,\,NN}]+V\,V_{,\,N}\,R_{,\,N}
      \right],\label{p2}
\end{equation}
respectively.

Equally, the temperature of the thermal bath  $T$ from Eq.(\ref{T}) results
\begin{equation}
  T=\left[\frac{R\,V_{,\,N}}{4C_\gamma\,(1+R)}\right]^{1/4}.\label{TTT}
\end{equation}
Here, we have used Eq.(\ref{dV}).

The relation between the e-folding $N$ and the scalar field can be written as
 \begin{equation}
   \int
   \left[\frac{V_{,\,N}}{V(1+R)}\right]^{1/2}\,dN=\int\,d\phi.\label{NF}
 \end{equation}

In this form, the scalar spectral index $n_S$ can be rewritten in
terms of the e-folds $N$, considering the Eqs.(\ref{nsG}),
(\ref{p1}) and (\ref{p2}) such that
$$
n_S-1=-\frac{(9R+17)}{8(1+R)}\,\frac{V_{,\,N}}{V}-\frac{(9R+1)}
{4(1+R)}\,\frac{\Gamma_{,\,N}}{\Gamma}
$$
\begin{equation}
+\frac{3}{4}\frac{1}{(1+R)}\,\frac{1}{VV_{,\,N}}\left[(1+R)[V_{,\,N}^2+V\,V_{,\,NN}]+V\,V_{,\,N}\,R_{,\,N}
      \right].
\end{equation}
For  the tensor to scalar ratio we have
\begin{equation}
  r=\frac{{\cal{P_T}}}{{\cal{P_S}}}=\frac{8\,V_{,\,N}}{(1+R)^{3/2}}\frac{1}{\sqrt{3V}\,T},\label{gr8}
\end{equation}
where the temperature of the thermal bath $T$ is given by
Eq.(\ref{TTT}). Here, we have considered  Eqs.(\ref{gr}) and
(\ref{p1}).

In the following, we will restrict ourselves to the weak
 and strong  dissipation regimes in order to   reconstruct under a general formalism  the model of warm
inflation from the cosmological parameters  $n_S(N)$ and $r(N)$.

\section{ The  weak dissipative regime.\label{section3a}}

We begin by considering the reconstruction for the case in which
the model of warm inflation evolves in the weak scenario in  which
the dissipation coefficient $\Gamma\ll3H$ (or equivalently
$R\ll1$). During this regime and considering Eq.(\ref{NF}),
 the relation between the e-folding $N$ and the scalar field, is
 given by
\begin{equation}
   \int
   \left[\frac{V_{,\,N}}{V}\right]^{1/2}\,dN=\int\,d\phi.\label{NF2}
 \end{equation}
On the other, from Eq.(\ref{nsG}) the spectral index $n_S$ when
$R\ll1$ becomes
\begin{equation}
n_S-1=-\frac{17}{4}\,\epsilon-\frac{1}
{4}\,\beta+\frac{3}{2}\,\eta,\label{nsGsd}
\end{equation}
and by considering the slow roll parameters given by
Eqs.(\ref{p1}) and (\ref{p2}), then the scalar spectral index in
this regime  can be rewritten as
\begin{equation}
n_S-1=\frac{1}{4}\left[-\frac{11\,V_{,\,N}}{2\,V}+3\frac{V_{,\,NN}}{V_{,\,N}}
-\frac{\Gamma_{,\,N}}{\Gamma}\right]=
\frac{1}{4}\left[\ln\left(\frac{V_{,\,N}\,^3}{V^{11/2}\,\,\Gamma}\right)\right]_{,N}\,.
\label{nsGsd2}
\end{equation}

From Eq.(\ref{gr}) the tensor to scalar ratio $r$ during this
scenario can be written as
\begin{equation}
r=8\left[\frac{4\sqrt{3}\,C_\gamma}{9}\,\frac{V_{,\,N}\,^3}{V^{3/2}\,\Gamma}\right]^{1/4}.\label{rw}
\end{equation}
Here, we have used that the temperature of the thermal bath is
given by $T=\left[\frac{\Gamma\,V_{,\,N}}{4\sqrt{3}\,C_\gamma
\,V^{1/2}}\right]^{1/4}$, during this scenario.

On the other hand, by combining  Eqs.(\ref{nsGsd2}) and (\ref{rw})
we find that the effective potential in terms of the e-folds $N$
can be written as
\begin{equation}
V(N)=V=\frac{r}{\tilde{C}_\gamma^{1/4}}\,\exp[\int(1-n_S)dN],\label{Vw}
\end{equation}
 and the coefficient
dissipation $\Gamma(N)$ becomes
\begin{equation}
\Gamma(N)=\Gamma=\frac{\tilde{C}_\gamma}{r^4}\,
\left[\frac{V_{,\,N}}{V^{1/2}}\right]^3=\left[\frac{\tilde{C}_\gamma^{1/4}}{r}\right]^{5/2}\,\left[\frac{r_{,\,N}}{r}+(1-n_S)\right]^{3}
\,\exp\left[\frac{3}{2}\int(1-n_s)dN\right],\label{GGd}
\end{equation}
where the constant $\tilde{C}_\gamma$ is defined as 
$\tilde{C}_\gamma=8^5\sqrt{3}C_\gamma/18$.

Here, we mention that the Eqs.(\ref{NF2}), (\ref{Vw}) and
(\ref{GGd}) are the fundamental equations in order to reconstruct
of the effective potential $V(\phi)$ and of the dissipation
coefficient $\Gamma(\phi)$, during this regime from the attractors
$n_S(N)$ and $r(N)$.

\section{ The  strong dissipative regime.\label{section3b}}

Now, we assume that the model of warm inflation evolves in the
strong dissipative regime in which the coefficient
 $\Gamma\gg3H$. Thus, from Eq.(\ref{NF}) the
relation between the e-folding $N$ and the scalar field during
this regime, is given by
\begin{equation}
   \int
   \left[\frac{V_{,\,N}}{V\,R}\right]^{1/2}\,dN=\int\left[\frac{\sqrt{3}\,V_{,\,N}}{\sqrt{V}\,\Gamma}\right]^{1/2}\,dN=\int\,d\phi.\label{NF3}
 \end{equation}

By considering  Eq.(\ref{nsG}) the spectral index in this regime
becomes
\begin{equation}
n_S-1=-\frac{9}{4R}\,\epsilon-\frac{9}
{4R}\,\beta+\frac{3}{2R}\,\eta,\label{nsGs}
\end{equation}
and taking into account  the slow roll parameters in this scenario
($R\gg1$), then the scalar spectral index $n_S$ can be rewritten
as

\begin{equation}
n_S-1=\frac{3}{4}\left[-\frac{V_{,\,N}}{2V}+\frac{V_{,\,NN}}{V_{,\,N}}+\frac{R_{,\,N}}{R}
-3\frac{\Gamma_{,\,N}}{\Gamma}\right]=
\frac{3}{4}\left[\ln\left(\frac{V_{,\,N}\,R}{V^{1/2}\,\Gamma^3}\right)\right]_{,N},
\label{nsGs2}
\end{equation}
o equivalently
\begin{equation}
n_S-1=
\frac{3}{4}\left[\ln\left(\frac{V_{,\,N}}{3^{1/2}\,V\,\Gamma^2}\right)\right]_{,N}.
\label{nsGs3}
\end{equation}
Here, we have considered Eqs.(\ref{inf2}) and (\ref{rG}).

During this regime, the tensor to scalar ratio $r$ results
\begin{equation}
r=\frac{8V_{,\,N}}{R^{3/2}}\,\frac{1}{\sqrt{3V}\,T}=\frac{1}{\Gamma^{3/2}}\,
\left[\bar{C}_\gamma\,V\,V_{,\,N}\,^3\right]^{1/4}.\label{rs}
\end{equation}
Here, we have used that the temperature of the thermal bath $T$ in
the strong regime is given by $T=(V_{,\,N}/4C_\gamma)^{1/4}$ and the constant
 $\bar{C}_\gamma$ is defined as $\bar{C}_\gamma=8^5\,3C_\gamma/2$.

By combining Eqs.(\ref{nsGs3}) and (\ref{rs})  we find 
that 
the effective potential $V(N)$ can be
written as
\begin{equation}
V(N)=\frac{r}{3^{3/8}\,\bar{C}_\gamma^{1/4}}\,\exp[\int(1-n_S)dN].\label{ps}
\end{equation}
Curiously, we find  
that the potentials $V(N)$ in the weak and  strong dissipative 
regime have the same structure, i.e., $V(N)\sim r \exp[\int(1-n_S)dN]$ , see 
Eqs. (\ref{Vw}) and (\ref{ps}).
 
The dissipative coefficient $\Gamma$ in terms of the e-folds $N$
can be expressed as
\begin{equation}
\Gamma(N)=\frac{\bar{C}_\gamma^{1/6}}{r^{2/3}}\,(V^{1/3}\,V_{,\,N})^{1/2}=\frac{1}{3^{1/4}}\,
\left[\frac{r_{,\,N}}{r}+(1-n_S)\right]^{1/2}\,\exp\left[\frac{2}{3}\int(1-n_s)\right]dN
.\label{Gs}
\end{equation}
Thus, we find that the dissipation coefficient given by  Eq.(\ref{Gs}) is different from the obtained
in the weak regime, see Eq.(\ref{GGd}). Also, we observe that the coefficient $\Gamma(N)$ 
does not depend of the constant $C_\gamma$, during the strong dissipative regime.

Again, we refer to that the Eqs.(\ref{NF3}), (\ref{ps}) and
(\ref{Gs}) are the fundamental  expressions  in order to
reconstruct of the effective potential $V(\phi)$ and
$\Gamma(\phi)$ from the quantities $n_S(N)$ and $r(N)$, during the
strong dissipative regime.

\section{ An example.\label{example}}

In this section we apply the formalism of above to the two
dissipative regimes (weak and strong) in the scenario of warm inflation, considering the simplest
example for the cosmological quantities  $n_S(N)$ and $r(N)$, in
order to reconstruct analytically the effective potential
$V(\phi)$ and dissipative coefficient $\Gamma(\phi)$. Following,
Refs.\cite{Staro,T,Chiba:2015zpa} we consider that the spectral
index is given by
\begin{equation}
n_S-1=-\frac{2}{N},\label{e1}
\end{equation}
and the tensor to scalar ratio as
\begin{equation}
r=\frac{1}{N(1+\xi N)},\label{e2}
\end{equation}
where $\xi$ corresponds to a constant. In this sense, if we
consider that in particular the number $N$ before the end of inflationary epoch at
the horizon exit corresponds to  $N \simeq 60$, then the  scalar
spectral index and the tensor to scalar ratio given by relations
(\ref{e1}) and (\ref{e2}) are well corroborate by observational
data if $\xi>-1/72$ \cite{Planck,Ob2}. In the
following we will assume that the number of e-folds $N$ is large,
for values of $N\sim \,\,\mathcal{O} (10^2)$.

As we mentioned before, the attractors given by Eqs.(\ref{e1}) and
(\ref{e2}) in the limit $\xi N\gg 1$ (such that $r\propto 1/N^2$)
 in the framework of cold inflation can be obtained in the
E-model\cite{E} and also in the model of the Higgs inflation with
the nonminimal coupling \cite{Higgs}, see also Ref.\cite{Higgs2}.
A generalization of the attractors $r(N)$ and $n_S(N)$  are given
by $r=12\sigma/N^2$ and $n_S-1=-2/N$ or also called $\sigma$
attractor (or usually called $\alpha$ attractor) which was
proposed in Ref.\cite{A1}, see also Ref.\cite{A2}.

\subsection{ The weak regime.\label{example1}}

By considering the spectral index given by Eq.(\ref{e1}) we find
that
$\exp[\int (1-n_S)]dN=\frac{N^2}{\alpha}$,
where the quantity $\alpha$ corresponds to the integration
constant. From Eq.(\ref{Vw}) we obtain that the effective
potential in terms of the e-folding becomes
\begin{equation}
V(N)=\frac{1}{\alpha\,\tilde{C}_\gamma^{1/4}}\left[\frac{1}{\xi+1/N}\right]\,.\label{potwe}
\end{equation}

Thus, we find that
$V_{,\,N}=\alpha^{-1}\tilde{C_\gamma}^{-1/4}(\xi N+1)^{-2}$, and
then the quantity $V_{,\,N}/V^2=\alpha \tilde{C_\gamma}/N^2$
suggests that the integration constant $\alpha>0$, since
 $\tilde{C_\gamma}$ and $V_{,\,N}$  are positives, see
Eq.(\ref{dV}). Also, we mention that the potential given by
Eq.(\ref{potwe}), is similar to that found in
Ref.\cite{Chiba:2015zpa} for cold inflation,  where the
reconstruction of $V(N)$ is only obtained from $n_S(N)$. In this sense, from the
potential (\ref{potwe}) we identify that
$\alpha\,\tilde{C_\gamma}^{1/4}$ corresponds to $\alpha$ and the
parameter $\xi\rightarrow \beta/\alpha$ from
cold inflation\cite{Chiba:2015zpa}.

By considering Eq.(\ref{GGd}), we find that the dissipation coefficient  in
terms of the number $N$ becomes
\begin{equation}
\Gamma(N)=\Gamma_0\,N^{5/2}\,(1+\xi
N)^{-1/2},\,\,\,\,\mbox{where}\,\,\,\,\Gamma_0=\frac{\tilde{C_\gamma}^{5/8}}{\alpha^{3/2}}.\label{G1}
\end{equation}

From Eqs.(\ref{potwe}) and (\ref{G1}) we find that the rate $R$ as
a function of the number of e-folds in this regime is given by
\begin{equation}
R(N)=\frac{\Gamma}{3H}\approx\frac{\Gamma}{\sqrt{3V}}=\frac{\tilde{C}_\gamma^{3/4}}{\sqrt{3}\;\alpha}\,N^2.\label{RR1}
\end{equation}
Here, we note that during the weak regime the ratio $R(N)$ does
not depend of the
 constant $\xi$, when it is expressed in terms of the number of e-folds $N$. Also, in order to obtain a scenario of
 weak dissipation in which $R\ll1$, we find a lower bound for
the parameter $\alpha$, given by
$\alpha\gg\frac{\tilde{C}_\gamma^{3/4}}{\sqrt{3}}\,N^2$. In
particular for large $N$ in which  $N=60$, we obtain the lower
limit $\alpha\gg8^3\times41.338\sim\,\mathcal{O}(10^7).$ Here, we have used
$C_\gamma=70$\cite{warm,warm2}.

In this form, during the stage of warm inflation we obtain a lower
bound for  the integration constant $\alpha$,  considering the
condition of the weak dissipative regime $\Gamma\ll3H$. Also, we
mention that this lower bound for the integration constant
$\alpha\gg\frac{\tilde{C}_\gamma^{3/4}}{\sqrt{3}}\,N^2$, can not
be obtained in the case of the reconstruction of the standard cold
inflation from the background level, and it is only possible to
say that $\alpha>0$ \cite{Chiba:2015zpa}.

 On the other hand, from Eq.(\ref{NF2}) we obtain
that the relation between $N$ and $\phi$ is given by the integral
\begin{equation}
\int\sqrt{\frac{1}{N(1+\xi N)}}\,dN=\int d\phi.\label{int2}
\end{equation}
Analogously as it occurs in the model of cold inflation,   this
integral depends on the sign of the constant $\xi$. In the case in
which $\xi>0$, we find that the integral given by Eq.(\ref{int2})
becomes
\begin{equation}
N=\xi^{-1}\,\sinh^2\left[\frac{\sqrt{\xi}}{2}(\phi-\phi_0)\right],\label{NWE1}
\end{equation}
where $\phi_0$ denotes an integration constant. In this form,
considering Eq.(\ref{NWE1}) the potential given by Eq.(\ref{potwe})
can be written in terms of the scalar field as
\begin{equation}
V(\phi)=V_0\,\tanh^2\left[\frac{\sqrt{\xi}}{2}(\phi-\phi_0)\right],\,\,
\,\,\,\,\mbox{where}\,\,\,\,\,V_0=\frac{1}{\alpha\,\xi\,\tilde{C}_\gamma^{1/4}}.\label{Vd1}
\end{equation}
This effective potential corresponds to the  potential  studied in
the T-model \cite{T}, see also Ref.\cite{Chiba:2015zpa} from the
reconstruction. From the point of view of the reconstruction, we noted that in
the case of the weak dissipative regime, the equation that relates
$N=N(\phi)$ (see Eq.(\ref{NF2})) is equivalent to the model of
cold inflation, since during the weak  dissipative regime ($3H\gg \Gamma$)
the expression to find $N=N(\phi)$ is the same, however, the initial 
fluctuations are of a different nature (thermal and quantum) and hence $n_S$ and 
the ratio $r$. In this sense, we notice that utilizing the specific  attractors  
for $n_S$ and $r$ given by Eqs.(\ref{e1}) and (\ref{e2}), the reconstruction in 
the model of warm inflation during the stage of weak dissipative regime 
coincides with the stage of cold inflation and it is a mere coincidence.
 Besides, we mention that this
potential can also adapted to provide the Starobinsky model and
$\alpha-$attractor model, see Ref.\cite{Chiba:2015zpa}.

 From Eqs.(\ref{G1}) and (\ref{NWE1}) we obtain that the
dissipation coefficient in terms of the scalar field is given by
\begin{equation}
\Gamma(\phi)=\frac{\Gamma_0}{\xi^{5/2}}\,\tanh\left[\frac{\sqrt{\xi}}{2}(\phi-\phi_0)\right]\,
\sinh^4\left[\frac{\sqrt{\xi}}{2}(\phi-\phi_0)\right].\label{Gd1}
\end{equation}
This suggests that, in order to obtain during the weak dissipative
regime the associated effective potential
 to the T-model or classes of inflationary models, with
an attractor point given by Eqs.(\ref{e1}) and (\ref{e2}), then
necessarily the dissipative coefficient should  be given by
expression (\ref{Gd1}). The dissipation coefficient for large $\sqrt{\xi}\phi$
can be approximated to
\begin{equation}
\Gamma(\phi)\simeq\frac{\Gamma_0}{16\xi^{5/2}}\,e^{2\sqrt{\xi}(\phi-\phi_0)}\,
\left(1-4e^{-\sqrt{\xi}(\phi-\phi_0)}\right)\sim
e^{2\sqrt{\xi}(\phi-\phi_0)} ,\label{Gdf1}
\end{equation}
and the effective potential (\ref{Vd1}) to the Starobinsky model or the
$\alpha-$attractor model \cite{Chiba:2015zpa}.

On the other hand,  an important situation occurs when the
integration constant $\xi$ is negative, since in particular for
$N=60$ and $\xi=-1/72$, the tensor to scalar ratio $r$ takes the
upper bound from Planck $r=0.1$\cite{Planck}. In this sense,
assuming  the case in which the constant $\xi<0$ and considering
that $\mid\xi^{-1}\mid>N$, then  the integration given by
Eq.(\ref{int2}) becomes
\begin{equation}
N=-\xi^{-1}\,\sin^2\left[\frac{\sqrt{-\xi}}{2}(\phi-\phi_0)\right],\label{NWE}
\end{equation}
and we find that the effective potential in terms of the scalar
field is similar to the found in Ref.\cite{Chiba:2015zpa} and becomes
\begin{equation}
V(\phi)=-V_0\,\tan^2\left[\frac{\sqrt{-\xi}}{2}(\phi-\phi_0)\right].\label{Vd2}
\end{equation}
However, the dissipative coefficient $\Gamma$ as a
function of the scalar field can be written as
\begin{equation}
\Gamma(\phi)=\frac{\Gamma_0}{(-\xi)^{5/2}}\,\tan\left[\frac{\sqrt{-\xi}}{2}(\phi-\phi_0)\right]\,
\sin^4\left[\frac{\sqrt{-\xi}}{2}(\phi-\phi_0)\right]. \label{Gd2}
\end{equation}

Finally in the situation in which the  constant $\xi=0$, the
relation between the number of e-folds and the scalar field
results $N=(\phi-\phi_1)^2/4$ and the potential is given by
\begin{equation}
V(\phi)=\frac{1}{4\alpha\,\tilde{C}_\gamma^{1/4}}\,(\phi-\phi_0)^2,\label{Vd3}
\end{equation}
corresponding to chaotic potential\cite{Chiba:2015zpa}.
  The dissipation coefficient as a function of
$\phi$ in this case becomes
\begin{equation}
\Gamma(\phi)=\frac{\Gamma_0}{32}\,(\phi-\phi_0)^5,\label{Gd3}
\end{equation}
with a dependency power-law in which $\Gamma\propto\phi^5$.
Here, we have used Eq.(\ref{G1}).

We emphasize that the reconstruction of the  effective potentials
given by Eqs.(\ref{Vd1}),(\ref{Vd2}) and (\ref{Vd3}) in the weak
dissipative regime are the same as those found in
Ref.\cite{Chiba:2015zpa} for the case of cold inflation only
assuming the scalar spectral index $n_S(N)=1-2/N$.
\begin{figure}[th]
\includegraphics[width=3.2in,angle=0,clip=true]{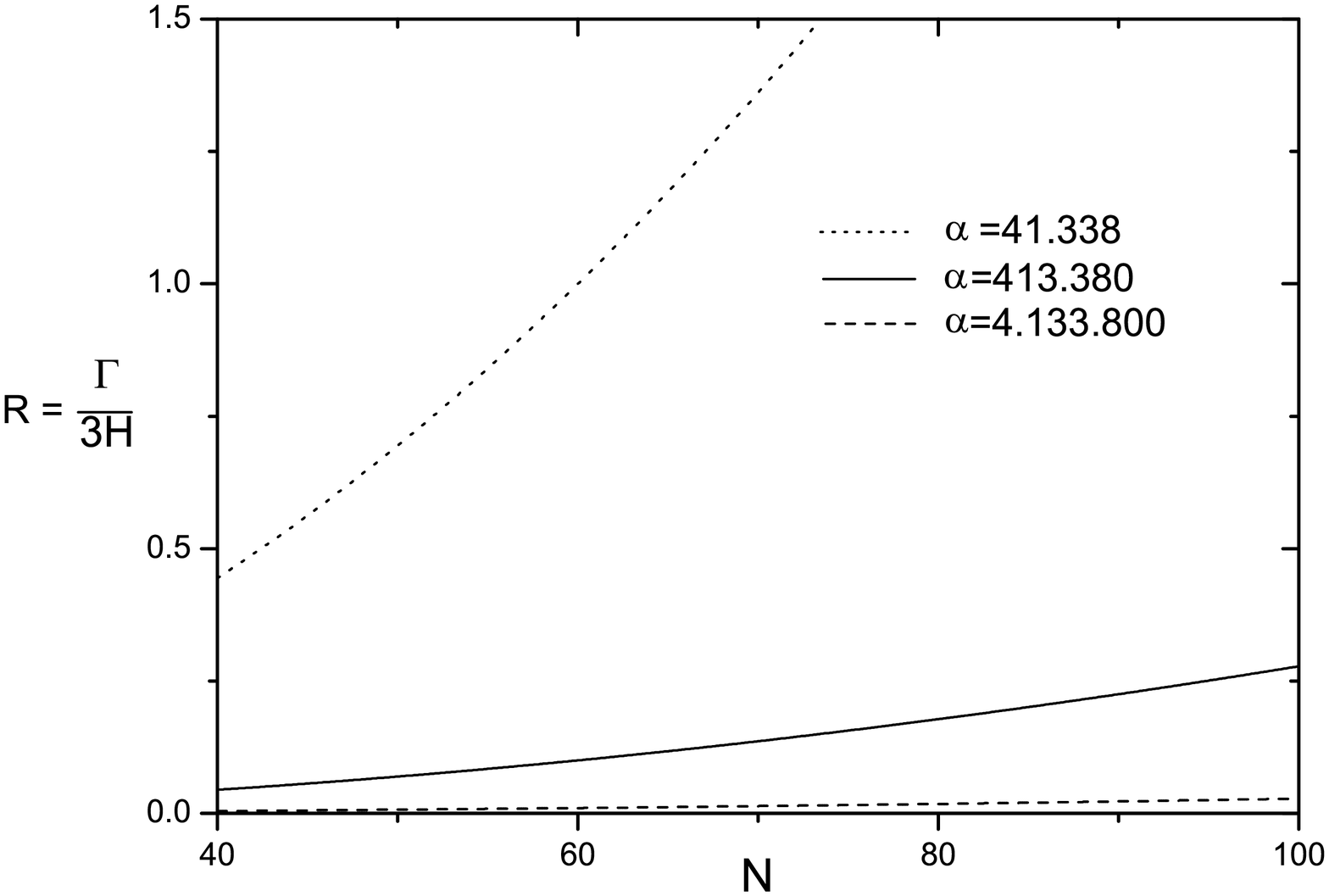}
\includegraphics[width=3.2in,angle=0,clip=true]{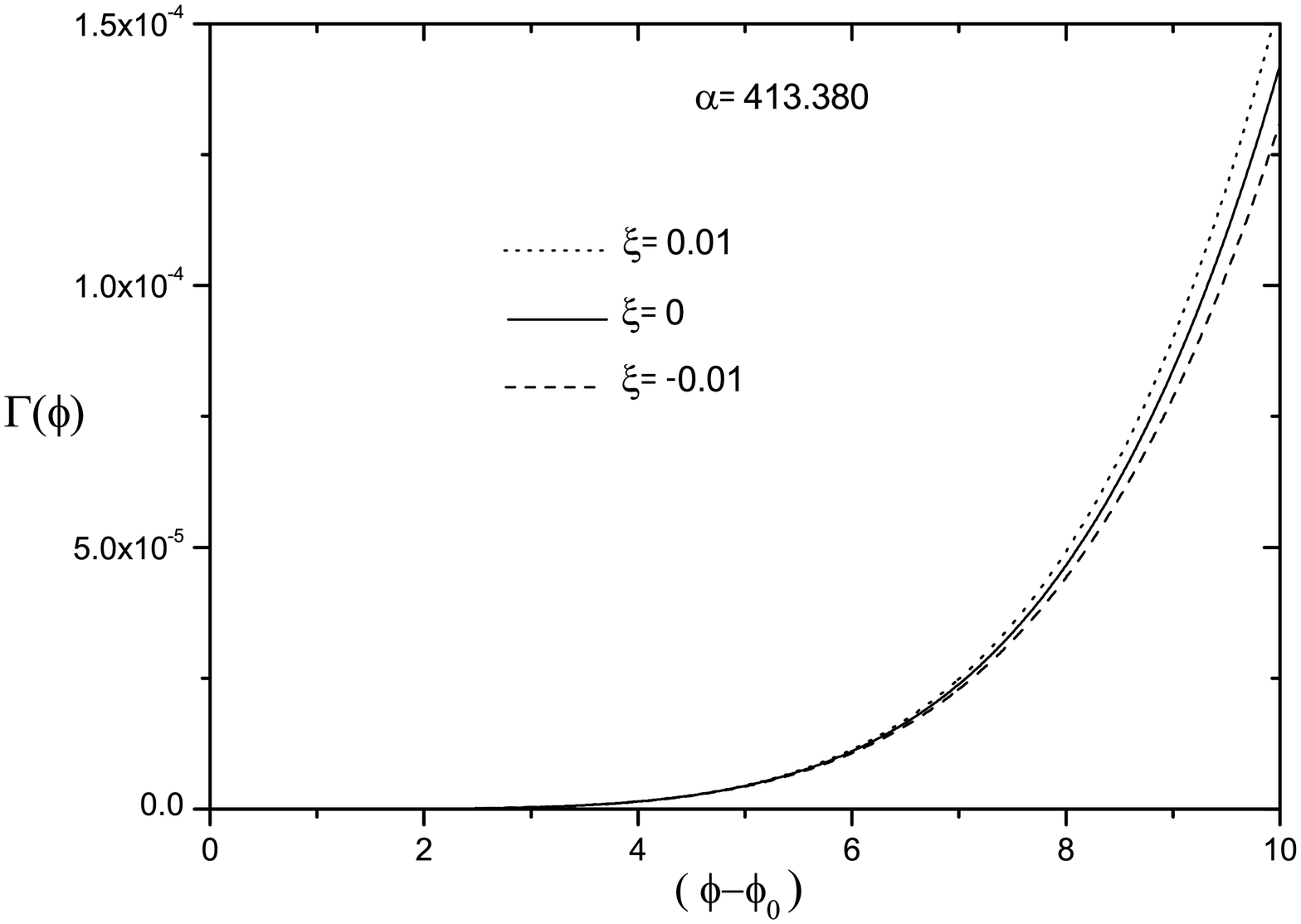}
{\vspace{-0.01 cm}\caption{The dependence of the ratio
$R=\frac{\Gamma}{3H}$ versus the number of e-folds $N$ (left
panel) and the dependence of the dissipation coefficient $\Gamma$
versus the scalar field (right panel) during the weak dissipative
regime. From Eq.(\ref{RR1}) we plot $R=R(N)$ in which the dotted
solid and dashed lines correspond to three different values of
$\alpha$ (left panel). From Eqs.(\ref{Gd1}), (\ref{Gd2}) and
(\ref{Gd3}) we plot $\Gamma=\Gamma(\phi)$ for three different
values of $\xi\gtreqqless 0$, in which we have fixed
$\alpha=413.380$ (right panel).
 Also, in these plots we have used $C_\gamma=70$.
 \label{fig1}}}
\end{figure}

In figure \ref{fig1} we show the evolution of the ratio
$R=\frac{\Gamma}{3H}$ versus the number of e-folds $N$ (left
panel) and the dependence of the dissipative coefficient $\Gamma$
on the scalar field (right panel) during the weak dissipative
regime $R\ll1$. The left panel shows the condition of the weak
dissipative regime in which $\Gamma\ll3H$, for three different
values of $\alpha$. In order to write down the rate $R=\Gamma/3H$
in terms of the e-folding $N$ during this regime, we consider
Eq.(\ref{RR1}). The right panel shows the evolution of the
dissipation coefficient $\Gamma$ as a function of the scalar
field. Also, in order to write down the coefficient $\Gamma$ in
terms of the scalar field, we consider Eqs.(\ref{Gd1}),
(\ref{Gd2}) and (\ref{Gd3}) for three different values of
$\xi\gtreqqless 0$, in which we have fixed
$8^{-3}\alpha=413.380\,\sim\mathcal{O}(10^6)$. In both panels we have
considered $C_\gamma=70$. From the left panel, we observe that the
condition for the weak dissipative regime ($R\ll1$) is satisfied
for the values of  the integration constant
$8^{-3}\alpha\gg41.338\sim\,\mathcal{O}(10^5)$. Also, we consider in
this plot the  limit case of the weak scenario in which the rate
$R=1$, corresponding  to the values $N=60$ and $8^{-3}\alpha=41.338$,
respectively (dotted line).

 From
right panel, we note that the behaviors of the different
parameters $\Gamma=\Gamma(\phi)$ for the   values of
$\xi\gtreqqless 0$ are similar. Also, we mention that from the
relation  given by Eq.(\ref{e2}) and considering  values  $\xi>0$ together
with  large $N$ i.e., $N\sim \,\,\mathcal{O} (10^2)$, the tensor
to scalar ratio $r\sim 0$.

\subsection{ The strong regime.\label{example2}}

By assuming the strong dissipative scenario in which $\Gamma\gg3H$  and from Eq.(\ref{ps}) we obtain that the effective
potential in terms of the e-folds $N$ results
\begin{equation}
V(N)=\frac{1}{3^{3/8}\alpha\,\bar{C}_\gamma^{1/4}}\left[\frac{1}{\xi+1/N}\right]\,,\label{potwe2}
\end{equation}
and this potential is similar to Eq.(\ref{potwe}) for the weak regime. We emphasize that the integration
constant $\alpha>0$. From Eq.(\ref{Gs}), 
we obtain that the dissipation coefficient in terms of the e-fonds is given by 
\begin{equation}
\Gamma(N)=\bar{\Gamma}_0\,N^{5/6}\,(1+\xi
N)^{-1/2},\,\,\,\,\mbox{where}\,\,\,\,\bar{\Gamma}_0=
\frac{1}{3^{1/4}\alpha^{2/3}}.\label{G12}
\end{equation}
As, we mentioned before this coefficient does not depend of the parameter 
$C_\gamma$ during the strong  regime.

Now, from Eqs.(\ref{ps}) and (\ref{Gs}) we obtain that the ratio $R$  during the strong dissipative regime is given 
by
\begin{equation}
R=\frac{\Gamma}{3H}\approx\frac{\Gamma}{\sqrt{3V}}=\frac{\bar{C_\gamma}^{1/8}}{3^{9/16}\alpha^{1/6}}\,
N^{1/3},\label{GN2}
\end{equation}
and this ratio does not depend of the parameter $\xi$, in analogy to the weak 
regime. From the condition of the strong regime in which $R\gg 1$,
we find an upper bound for the parameter $\alpha$ given by
 $\frac{\bar{C_\gamma}^{3/4}N^2}{3^{27/8}}\gg \alpha$. For the case in which 
 $N=60$, we obtain that the upper limit for $\alpha$ is given by  
 $7\times 10^{6}\sim \,\,\mathcal{O} (10^7)\gg\alpha$. 
 
Nevertheless, from these solutions  we
find a transcendental equation from Eq.(\ref{NF3}) to express the
number of e-folds in function of the scalar field. Hence, in order
to obtain analytical expressions  for $V(\phi)$ and $\Gamma(\phi)$
and therefore the reconstructions, we can study  the potential and 
the dissipation coefficient  in the
limits $ N\gg 1/\xi$ and  $ N\ll 1/\xi$.

We start with the limit $\xi N\gg 1$. For large N and in
particular for $N=60$,  we find a lower bound for the constant
$\xi$  given by $\xi\gg 1/60\simeq 0.017$. On the other hand,  the
tensor to scalar ratio $r(N)$ given by Eq.(\ref{e2}) in this limit is
approximately
\begin{equation}
r(N)=\frac{1}{N(1+\xi N)}\approx\frac{1}{\xi\,N^2},\label{aa1}
\end{equation}
with $\xi$ a positive quantity. Here, we note that the attractor
given by Eq.(\ref{aa1}) corresponds to the $\sigma-$attractor, in
which $\xi=(12\sigma)^{-1}$ \cite{A1,A2} or also to  $T$-model
when $\xi$ takes the value $\xi=1/12$ in cold
inflation\cite{T}.

 In this context,
in which  $\xi N\gg 1$, the effective potential $V(N)$ given by
Eq.(\ref{potwe2})  results
\begin{equation}
V(N)\approx\frac{1}{3^{3/8}\alpha\,\bar{C}_\gamma^{1/4}\xi}=Cte.
\end{equation}
Thus, the universe presents an exponential expansion, 
since $H\propto V^{1/2}=$ constant and from Eq.(\ref{NF3}) 
we find  $\phi=\phi_0=Cte.$, because $d\phi/dN=0$.  This suggests that the reconstruction does not work
during the strong regime when $r\propto N^{-2}$ and $n_S-1\propto N^{-1}$.

On the other hand, now we consider the limit in which
  $\xi N\ll 1$ where $r(N)\approx  1/N$. We note this the attractor for large $N$
  and in particular
for $N=60$ results $r(N=60)\approx 1/60\simeq
0.02$, wherewith still  this attractor is well supported by the Planck data.

  From Eq.(\ref{potwe2}) we find that
the effective potential $V(N)$ becomes
\begin{equation}
V(N)\approx\,\frac{1}{3^{3/8}\alpha\,\bar{C}_\gamma^{1/4}}\,N.\label{V5}
\end{equation}
Considering that the integration constant $\alpha\ll 7\times10^{6}$, we find a 
lower bound for the effective potential for large $N$
 (in particular $N=60$) given by $V(N=60)\gg10^{-7}$ (in units of 
 $m_p^4$, with $m_p$ the Planck mass).

 The dissipation coefficient from Eq.(\ref{G12}) is given by
\begin{equation}
\Gamma(N)\approx\,\bar{\Gamma}_0\,N^{5/6}
\,\,.\label{G5}
\end{equation}
Here, we observe that the potential $V(N)$ and the dissipation
coefficient $\Gamma(N)$ do not depend on the parameter $\xi$,
since $r(N)\approx 1/N$.

From Eq.(\ref{NF3}) we find that the relation between the number
of e-folds and the scalar field in the limit $\xi N\ll 1$ can be written as
\begin{equation}
\frac{1}{\tilde{\alpha}}\int N^{-2/3}dN=\int 
d\phi,\,\,\,\,\mbox{where}\,\,\,\,\tilde{\alpha}=\frac
{\bar{C}_\gamma^{1/16}}{3^{9/32}\,\alpha^{1/12}}.
\end{equation}
Integrating we have
\begin{equation}
N\approx N_0\,(\phi-\phi_0)^3,\label{N1}
\end{equation}
where $N_0=(\tilde{\alpha}/3)^3$
and $\phi_0$ corresponds to  an integration constant. Thus, the
reconstruction of the effective potential as a function of the
scalar field results
\begin{equation}
V(\phi)\approx\,V_0\,(\phi-\phi_0)^3,\,\,\,\,\mbox{where}\,\,\,\,
V_0=\frac{N_0}{3^{3/8}\alpha\bar{C}_\gamma^{1/4}}.
\end{equation}
Therefore, in this limit ($\xi N\ll 1$) the effective potential
corresponds to a cubic  potential.

Similarly, the
dissipation coefficient from Eqs.(\ref{V5}) and (\ref{N1}) in this
limit becomes
\begin{equation}
\Gamma(\phi)\approx
\Gamma_0\,(\phi-\phi_0)^{5/2},\,\,\,\,\mbox{where}\,\,\,
\,\,\Gamma_0=\bar{\Gamma}_0\,N_0^{5/6}\,\,,\label{Ga5}
\end{equation}
resulting in  a power law dissipative coefficient in which $\Gamma\propto \phi^{5/2}$.

In this sense, we observed that considering the attractor
$r(N)\approx 1/N$ (together with $n_S-1=-2/N$),  the effective
potential and the dissipation coefficient
 present a power law behavior during the strong regime, and
 its dependencies  with the scalar field (reconstruction) are given by  $V(\phi)\sim
 \phi^3$ and $\Gamma(\phi)\sim \phi^{5/2}$, respectively.

\section{Conclusions \label{conclu}}

In this paper we have studied the reconstruction from recent
cosmological observations in the framework of the warm inflation.
Under a general formalism of reconstruction, we have found
expressions for the effective potential and dissipative
coefficient
  in the context of the slow roll
approximation, motivated by the cosmological observations of the
scalar spectral index $n_S$ and tensor to scalar ratio $r$. In
this general analysis we have obtained from the cosmological
quantities $n_S(N)$ and $r(N)$ (where $N$ corresponds to the
number of e-folds), integrable expressions
 for the effective potential and dissipative coefficient. For warm inflation and 
 its reconstruction,   we have considered two different regimes, called the weak and strong dissipative
 regimes.

As a concrete example and in order to obtain the reconstructions
for the effective potential $V(\phi)$ and dissipation coefficient
$\Gamma(\phi)$, we have considered the attractors $n_S-1=-2/N$ and
$r=(N[1+\xi N])^{-1}$. Here,
 we have applied  our general results
considering   the  weak and strong dissipative regimes for these
attractors.

For the weak regime in which $\Gamma\ll3H$ (or equivalently $R \ll
1$) and considering the example or the attractors given by
Eqs.(\ref{e1}) and (\ref{e2}), we have obtained a lower bound for
the integration constant $\alpha$
 given by $\alpha\gg\frac{\tilde{C}_\gamma^{3/4}}{\sqrt{3}}\,N^2$, from the condition of weak dissipative regime i.e.,
   $R(N)\ll1$. In particular for the case in which  $N=60$ (large $N$), we have found
that the lower bound for the integration constant $\alpha$ given by 
  $\alpha\gg2\times 10^{7}\sim \mathcal{O}(10^7)$.
Also, we have obtained that during the weak regime the
reconstruction on the effective potentials are given by
Eqs.(\ref{Vd1}), (\ref{Vd2}) and (\ref{Vd3}), and it
 coincides with the obtained in the case of cold inflation\cite{Chiba:2015zpa}.
 Similarly,
 we have obtained that  the construction of the dissipative
 coefficients $\Gamma(\phi)$ depends on the sign of the parameter $\xi\gtreqless 0$.
 In particular for the case $\xi=0$ where the potential corresponds to the chaotic potential, we have found that the
 dissipative coefficient $\Gamma(\phi)\propto \phi^5$.

 For the case of the
 strong dissipative regime ($R\gg1$) we have obtained the potential
 and dissipative coefficient in terms of the number of e-folds. During this 
 regime, we have found that the potential $V(N)$ has the same structure that in 
 the weak regime.
 However, we could not find analytical solutions in order to
 obtain the number of e-fold  in terms of the scalar field in form to obtain the reconstruction of $V(\phi)$ and
 $\Gamma(\phi)$. In this sense, we have analyzed  
  the potential and the dissipation coefficient  in the
limits $ N\gg 1/\xi$ and  $ N\ll 1/\xi$, in order to obtain analytical solutions. 
In the case in which $r\propto N^{-2}$ (limit $N\gg 1/\xi$), we have obtained that 
the potential $V(N)=$ constant, and the reconstruction does not work. For the 
case in which $r\propto N^{-1}$ (limit $N\ll 1/\xi$) we have obtained that the 
potential and the dissipative coefficient in terms of the scalar field are given 
by $V(\phi)\propto \phi^3$ and $\Gamma(\phi)\propto \phi^{5/2}$, respectively.

Finally in this paper, we have not addressed the reconstruction of warm inflation in which
the effective potential and dissipative coefficient also depend of
the temperature of the thermal bath $T$, i.e.,  $V(\phi,T)$ and
$\Gamma(\phi,T)$\cite{taylorberera,w1,w3,w4,Moss:2008yb}. We hope
to return to this point in the near future.

\begin{acknowledgments}
 The author thanks Prof. $\emptyset$yvind  Gr$\emptyset$n by the comment on the 
 tensor to scalar ratio. 
 This work was supported
by PUENTE Grant DI-PUCV N$_{0}$ 123.748/2017.
\end{acknowledgments}


\end{document}